# Heavily Sr-Doped $La_2SrNi_2O_{7-\delta}$ as a Tetragonal Ruddlesden-Popper Phase at Ambient Pressure

Yuhang Zhang[1†], Xue Ming[1†], Cui-Qun Chen[2†], Wei Chen[3†], Tian-Yi Li[1], Zhe-Ning Xiang[1], Qing Li[1], Bing-hui Ge[3*], Dao-Xin Yao[2*], Xiyu Zhu[1]* and Hai-Hu Wen[1]*

[1] National Laboratory of Solid State Microstructures, Department of Physics, Center for Superconducting Physics and Materials, Collaborative Innovation Center of Advanced Microstructures, Nanjing University, Nanjing 210093, China

[2] Institute of Neutron Science and Technology, Guangdong Provincial Key Laboratory of Magnetoelectric Physics and Devices, State Key Laboratory of Optoelectronic Materials and Technologies, School of Physics, Sun Yat-sen University, Guangzhou 510275, China

[3] National Key Laboratory of Opto-Electronic Information Acquisition and Protection Technology, Leibniz International Joint Research Center of Materials Sciences of Anhui Province, Institutes of Physical Science and Information Technology, Anhui University, Hefei, 230601, China

[†] These authors have equal contributions. X. M. is on leave from School of Physics and Optoelectronic Engineering, Shandong University of Technology, Zibo 255000, China

* Corresponding authors with e-mails: bhge@ahu.edu.cn, yaodaox@mail.sysu.edu.cn, zhuxiyu@nju.edu.cn, hhwen@nju.edu.cn

**Abstract:** High-temperature superconductivity has been found in bilayer Ruddlesden-Popper (RP) nickelates in bulk samples under high pressure, or in thin films via compressive strain. In the superconducting state, a tetragonal structure with a straight Ni−O−Ni bond along *c*-axis has been commonly observed, together with the suppression or diminishing of the density-wave orders. Therefore, it remains an open question whether these factors are sufficient for achieving superconductivity at ambient pressure. Here we report the first successful synthesis of heavily Sr-doped $La_2SrNi_2O_{7-\delta}$ under high-pressure and high-temperature (HPHT) conditions with a flux method. X-ray diffraction and scanning transmission electron microscopy (STEM) confirm that the material adopts a tetragonal (*I*4/*mmm*) structure with an 180° Ni−O−Ni bond angle along $c$-axis. Resistance measurements reveal metallic behavior with a low-temperature upturn and no density-wave features are observed. However, neither pressure nor oxygen variation induces superconductivity. Density functional theory calculations indicate that the holes introduced by Sr doping are predominantly doped into the Ni-$3d_{z^2}$ orbital, leading to the appearance of a very large γ pocket on the Fermi surface at ambient pressure and significantly reducing the occupation of the Ni-$3d_{z^2}$ orbital. Combining the experimental observations with theoretical calculations, we attribute the absence of superconductivity to the serious deviation from the half-filling state of the Ni-$3d_{z^2}$ band, which is crucial for the interlayer antiferromagnetic interaction and thus for pairing. Our work unravels important issues for achieving superconductivity in bilayer nickelate system.

## 1. Introduction

Nickelate superconductivity has attracted widespread interest since the discovery of superconductivity in infinite-layer compounds[1]. Another breakthrough came in 2023, when Ruddlesden-Popper (RP) $La_3Ni_2O_7$ single crystals were found to exhibit superconductivity with a transition temperature ($T_c$) above 77 K under high pressure[2]. The rapid progress on the nickelate superconductivity has been well documented in several recent overview papers[3–6]. Extensive efforts have been focused on La-site doping with other rare-earth ions in $La_3Ni_2O_7$, leading to bulk superconductivity in $La_2PrNi_2O_{7-\delta}$ and a record $T_c$ of 96 K in $La_{3-x}Sm_xNi_2O_{7-\delta}$[7,8]. Moreover, superconductivity has also been observed in trilayer ($La_4Ni_3O_{10}$ and $Pr_4Ni_3O_{10}$) and hybrid ($La_2NiO_4{\cdot}La_3Ni_2O_7$ and $La_2NiO_4{\cdot}La_4Ni_3O_{10}$) bulk nickelates[9–16]. Most notably, ambient-pressure superconductivity with the highest onset transition temperature reaching ~60 K has recently been realized in compressively strained bilayer nickelate thin films[17–22]. These advances have greatly motivated efforts to elucidate the key factors governing the superconductivity and to realize possible ambient-pressure superconductivity in bulk samples.

At ambient pressure, $La_3Ni_2O_7$ crystallizes in an orthorhombic structure featuring two distorted vertex-sharing $NiO_6$ octahedra along $c$-axis, which are separated by an inner apical oxygen of the single layer LaO; each bilayer structure containing double $NiO_2$ planes are separated by the so-called rock-salt $La_2O_2$ layers[2,7,8]. Meanwhile, a density-wave (DW) transition was observed near 150 K[23–25]. Upon applying pressure, $La_3Ni_2O_7$ undergoes a structural transition from an orthorhombic phase (*Amam*) to a tetragonal phase (*I*4/*mmm*) with an intermediate orthorhombic *Fmmm* phase at low pressures[26,27], which is accompanied by the suppression of the DW order, and eventually superconductivity emerges[2,7,26–29]. In the trilayer compounds, a similar pressure-induced structural transition was observed and superconductivity occurs almost coincidently with the structure transition[9–11,30]. Moreover, superconductivity was also observed in compressively strained $La_3Ni_2O_7$ and $La_{3-x}Pr_xNi_2O_7$ thin films at ambient pressure, now the films show already a tetragonal phase[18–20]. Thus, the tetragonal phase with a straight Ni−O−Ni bond angle along *c*-axis is thought to be critical for enhancing interlayer coupling and inducing superconductivity[2]. However, some previous studies revealed no evidence of superconductivity in tetragonal $La_3Ni_2O_{7-\delta}$ and $La_4Ni_3O_{10}$, suggesting that the DW state may be a prerequisite for superconductivity[31,32]. Overall, it remains essential to provide a viable route to ambient-pressure superconductivity in bulk nickelates. One promising direction is to stabilize the tetragonal phase at ambient pressure through strategies such as La-site doping, combined with band-structure engineering, to explore the sufficient conditions for achieving superconductivity[33,34].

Aliovalent cation doping offers a powerful strategy to introduce charge carriers, modify the electronic structure, and thereby tune superconductivity and related properties[35,36]. In $La_3Ni_2O_7$, the antiferromagnetic (AF) super-exchange between Ni atoms of the double layers seems to be strong, this is mediated by the vertical hopping term $t_\perp$ and the apical O-2$p_z$ orbitals. The double layer structure of $NiO_2$ planes splits the Ni-3$d_{z^2}$ band into bonding and antibonding orbitals, with the former lying probably below the Fermi level ($E_F$)[2]. It has been proposed that the metallization of the Ni-3$d_{z^2}$ bonding band under high pressure, leading to the so-called γ Fermi pocket, may be essential for superconductivity[37–43]. Indeed, recent calculations have investigated the effects of carrier doping on the band structure and superconductivity in $La_3Ni_2O_7$, predicting that hole doping could raise the γ band and lead to a hole-like Fermi pocket near the Fermi energy (metallization of the σ-bonding bands) at ambient pressure[44–48]. Experimentally, however, Sr doping in bulk bilayer nickelates has proven to be challenging. In $La_{3-x}Sr_xNi_2O_7$ single crystals, it is only possible to incorporate Sr up to a concentration of x = 0.2 (even under 20 GPa), whereas polycrystals annealed under a high-pressure oxygen atmosphere allows only a Sr concentration of about $x < 0.1$, which is insufficient to substantially modify the band structure[49,50]. Therefore, high-quality hole-doped bulk nickelates with a tetragonal phase and a straight Ni−O−Ni bond angle along *c*-axis are urgently needed to clarify the interplay between band engineering, interlayer coupling, and superconductivity.

In this paper, we report the first successful synthesis of heavily Sr-doped $La_2SrNi_2O_{7-\delta}$ single crystals with the high pressure and high temperature (HPHT) technique, the samples adopt a tetragonal *I*4/*mmm* structure with a 180° Ni−O−Ni bond angle at ambient pressure. The samples exhibit a metallic behavior with a low-temperature upturn and no signatures of DW orders. However, neither pressure nor oxygen variation induces superconductivity. We attribute the absence of superconductivity to the serious deviation from half-filling of the Ni-$3d_{z^2}$ band, which pushes the γ pocket well above the Fermi energy and thereby weakens the interlayer antiferromagnetic coupling and pairing.

## 2. Results and discussion

### Structural and compositional characterizations of tetragonal $La_2SrNi_2O_{7-\delta}$

The $La_2SrNi_2O_{7-\delta}$ single crystals were synthesized using a flux method under HPHT conditions. In this work, $La_2SrNi_2O_{7-\delta}$ represents the nominal composition. The details of crystal growth process are provided in the Experimental Section. Single-crystal X-ray diffraction (SXRD) measurements are performed on the as grown $La_2SrNi_2O_{7-\delta}$ crystal to identify the ambient-pressure structure. Figure 1a displays the crystal structure of $La_2SrNi_2O_{7-\delta}$, crystallizing in ($n$=2) RP phase. The detailed crystallographic data of $La_2SrNi_2O_{7-\delta}$ are listed in Table 1. The SXRD data reveals a tetragonal structure with a space group *I4/mmm* (No. 139). This body-centered tetragonal symmetry is identical to that in bilayer nickelates at superconducting state under a high pressure, and is higher than the orthorhombic *Amam* (No. 63) structure observed in $La_3Ni_2O_{7-\delta}$ at ambient pressure[7,8,26,27]. Recent theoretical calculations also identify the suppression of orthorhombicity as a key ingredient for superconductivity[51].

Powder X-ray diffraction (PXRD) was also performed by grinding the as-grown single crystals into powder. Figure 1b shows Rietveld refinement profile of $La_2SrNi_2O_{7-\delta}$ samples. High purity of $La_2SrNi_2O_{7-\delta}$ phase is confirmed by indexing all the observed diffraction peaks, which shows remarkable consistency with the tetragonal *I*4/*mmm* model. The inset of Figure 1b illustrates that the Ni−O−Ni bond angle along *c* axis increases from 167.6° in $La_3Ni_2O_{7-\delta}$ to 180° in $La_2SrNi_2O_{7-\delta}$. Therefore, heavy Sr doping serves as an effective approach to suppress the structural distortion of the $NiO_6$ octahedron. This suggests that substitution on La site with alkaline earth elements provides a potential route to stabilize the tetragonal bilayer nickelate. We performed a more detailed structural analysis on PXRD data of $La_2SrNi_2O_{7-\delta}$ (red) and $La_2PrNi_2O_{7-\delta}$ (blue), as shown in Figure 1c. And the diffraction peaks were both well indexed with the orthorhombic and tetragonal structures, respectively. Notably, distinct peak structures merge in the Sr doped samples, for example the merging of the (020) and (200) peaks at ~33.1°, the (026) and (206) peaks at ~ 42.8°, and the (135) and (315) peaks at ~58.6°. This comparative observation further confirms that heavy Sr doping induces a structural transition towards a higher symmetry. In addition, a pronounced rightward shift of the (00$\underline{10}$) reflection is observed, indicating a significant compression of the *c*-axis lattice parameter in $La_2SrNi_2O_{7-\delta}$.

The inset of Figure 1d shows a scanning electron microscopy (SEM) image of $La_2SrNi_2O_{7-\delta}$ microcrystals in which a single crystal with a length of about 30 micrometers can be easily visualized. Prior to imaging, the crystals were sequentially washed with deionized water and anhydrous ethanol in multiple cycles to remove residual flux. The observed rectangular edge morphology and smooth surface indicate high quality of these crystals. In contrast to the single crystals of other RP phases grown from flux, the $La_2SrNi_2O_{7-\delta}$ single crystals exhibit a significantly smaller thickness. As presented in Figure 1d, if normalizing the Ni content to 2, the EDS analysis yields a La: Sr: Ni atomic ratio of 1.86: 1.14: 2.02, which is close to the stoichiometry of $La_2SrNi_2O_{7-\delta}$.

**Microstructure Visualization and Symmetry confirmation**

As mentioned above, both SXRD and PXRD results show that our sample $La_2SrNi_2O_{7-\delta}$ crystallizes in a tetragonal phase at ambient pressure. To further characterize its microstructure, we performed high-angle annular dark-field (HAADF) scanning transmission electron microscopy (STEM) on the $La_2SrNi_2O_{7-\delta}$ single crystals. Figure 2a presents a large-area HAADF-STEM image taken along the [100] axis, where the positions of La/Sr, Ni atoms and $NiO_6$ octahedra were marked with colored circles and grey squares, revealing the well-ordered alternating bilayer stacks. In contrast to the octahedral tilting and distortion observed in $La_3Ni_2O_{7-\delta}$[52,53], the bilayer $NiO_6$ octahedra in $La_2SrNi_2O_{7-\delta}$ adopt a linear, unbuckled arrangement within each RP bilayer. Furthermore, selected area electron diffraction (SAED) was carried out to verify the tetragonal *I*4/*mmm* structure. As displayed in Figure 2b, the SAED pattern along the [001] axis exhibits a characteristic fourfold rotational symmetry, directly corroborating the tetragonal lattice ($a = b$) and unambiguously distinguishes it from the orthorhombic structure observed in $La_3Ni_2O_{7-\delta}$[52]. Notably, all observed ($hk0$) reflections satisfy the condition $h + k = 2n$, indicative of the body-centered *I*4/*mmm* symmetry. Figure 2c shows the SAED pattern along the [100] axis, which clearly reveals $b \neq c$, while the interaxial angle remains 90°, in agreement with tetragonal symmetry. In addition, all ($0kl$) reflections in Fig.2c obey the condition $k + l = 2n$. Therefore, all reflections are well indexed and fully comply with the systematic reflection condition for the *I*4/*mmm* space group, i.e., for *hkl*: $h + k + l = 2$n, consistent with the results from SXRD. The HAADF-EDS mapping results of $La_2SrNi_2O_{7-\delta}$ single crystals are presented in Fig. 2d, revealing the spatial distributions of La, Sr and Ni. Well-defined Ni blocks are clearly visible, manifesting the tetragonal bilayer stacking. Evidently, Sr preferentially occupies the La sites in the rock-salt layer $La_2O_2$, with only a small amount of Sr observed in the inner single layer LaO. This Sr site preference is consistent with the crystal structure shown in Figure 1a and analogous to observations in Pr-doped and Sm-doped bilayer nickelates[8,53]. Oxygen vacancy commonly exists in bilayer nickelates[21,54]. To clarify the oxygen vacancy distribution in our samples, we performed integrated differential phase contrast (iDPC) imaging on $La_3Ni_2O_{7-\delta}$ single crystals (Extended Data Fig.4). The oxygen vacancies in our samples are substantially reduced, which we attribute to the high oxygen partial pressure applied in the growth process.

**Transport and Magnetic Properties at ambient pressure**

To investigate the electrical transport and magnetic properties of $La_2SrNi_2O_{7-\delta}$ at ambient pressure, we conduct the temperature-dependent resistance measurements and the dc magnetization measurements. Fig. 3a shows the normalized resistivity of polycrystalline pellet of $La_2SrNi_2O_{7-\delta}$ crushed from single crystals and annealed at 500 °C in air (S1) and under one atmosphere of oxygen flow (S2). Both samples exhibit metallic behavior with a low-temperature upturn at approximately 88 K (S1) and 68 K (S2), respectively. The different upturn onset temperatures of the polycrystalline samples could be attributed to grain-boundary scattering. Notably, a much weaker low-temperature upturn was also observed in the as-grown single crystals measured at low pressures (Figure 3d). Combined with the iDPC imaging results (Extended Data Fig.4), the slight oxygen vacancies in quite few regions may lead to this weak resistance upturn in the low temperature for the single crystal samples [55,56]. Figure 3b shows the curves of temperature-dependent susceptibility of $La_2SrNi_2O_{7-\delta}$ measured at 5 kOe in zero-field-cooled (ZFC) and field-cooled (FC) modes. Over the entire temperature range from 2 K to 300 K, the magnetic susceptibility shows a Curie-Weiss (CW) like paramagnetic behavior and is fitted well with the CW law: $\boldsymbol{\chi(T) = \chi_0 + \frac{C}{T+T_\theta}}$ , where $\boldsymbol{C = \frac{N\mu_0\mu_{eff}^2}{3k_B}}$ is the Curie constant, $\chi_0$ denotes the temperature-independent term, and $T_\theta$ stands for the Curie-Weiss temperature. Therefore, with the fitted Curie constant $C$ = 0.137 emu K mol$^{-1}$ Oe$^{-1}$, the effective magnetic moment per formula of $La_2SrNi_2O_{7-\delta}$ is derived to be $\boldsymbol{\mu_{eff} = 1.05\ \mu_B}$. Neither the electrical resistance nor the magnetization shows any distinct density waves behavior.

**Search for Superconductivity via Oxygen/Ozone annealing and Pressure Tuning**

Tuning the oxygen content and applying external pressure are common approaches to search for superconductivity in the RP nickelates[2,17,18,57–59]. The modulation of oxygen content may alter the valence state of Ni in $La_2SrNi_2O_{7-\delta}$, potentially adjusting it back to the +2.5 state as found in $La_3Ni_2O_7$. However, the absence of internal apex oxygen under such conditions may still suppress superconductivity. Fig. 3c presents the transport data of samples subjected to different oxygen modification treatments: oxygen reduction with La metal in a sealed quartz tube (S3), and subsequently annealed with ozone treatment at 325 °C for 1 h (S4) and 3 h (S5), respectively. Details are presented in Experimental Section. XRD results in Extended Data Fig.3 indicate that while all these samples retain the original tetragonal structure, their transport behaviors differ markedly, underscoring the significant impact of the internal apex oxygen on transport. As we can see, sample S3 exhibits strong insulating behavior, which is associated with the severe deficiency of internal apex oxygen. After ozone annealing, the magnitude of $\rho(T)$ decreases substantially, and sample S5 even shows metallic behavior above 210 K. Although modifying the oxygen content can push the Ni electronic configuration more closer to the $3d^{7.5}$ state, the resulting loss of internal apex oxygen also suppresses the interlayer antiferromagnetic super-exchange interaction, thereby hindering superconductivity[54,60]. To explore the evolution of the electrical transport behavior and the possible emergence of superconductivity under pressure, we preformed high-pressure electrical resistance measurements on the as-grown $La_2SrNi_2O_{7-\delta}$ single crystals up to 51.9 GPa with the diamond anvil cells (DAC). As illustrated in Figure 3d, the resistance at low pressures exhibits metallic behavior accompanied by a weak resistance upturn in the low temperature region, and no DW like transition is observed. Although the upturn is gradually suppressed with increasing pressure, it persists up to the highest pressure and no superconductivity is observed. Combining the theoretical calculations presented later, we attribute the absence of superconductivity in the current samples to the strong deviation from the half-filling state of the Ni-$3d_{z^2}$ band caused by the heavy Sr doping, although the correlated γ pocket to lie well above the Fermi energy.

## DFT calculations of tetragonal $La_2SrNi_2O_7$

To further understand the electronic properties of $La_2SrNi_2O_7$, we performed density functional theory (DFT) calculations at ambient pressure and 25 GPa. Under the crystal field of $NiO_6$ octahedra, the Ni $3d$ orbitals split into $e_g$ and $t_{2g}$ sectors. The electronic bands from -1.5 eV to 3 eV around the Fermi level are predominantly of Ni-$e_g$ character (Fig. 4a), similar to the parent compound $La_3Ni_2O_7$[2]. The interlayer Ni σ-bond coupling mediated by the apical oxygen atoms gives rise to bonding and antibonding bands near the Fermi level. Notably, the substitution of $Sr^{2+}$ introduces one hole into the Ni-$e_g$ orbitals. DFT results show that the holes are mainly doped into the Ni-$3d_{z^2}$ orbital, driving the bonding band across the Fermi level at ambient pressure and resulting in the metallization of the σ-band. Figure 4c displays the ambient-pressure Fermi surface, which consists of two electron pockets (α and β) and one hole pocket (γ). Compared with the undoped $La_3Ni_2O_7$, the γ pocket appears at ambient pressure and is significantly enlarged because of the additional hole doping. This enlargement originates from the aliovalent Sr doping, which raises the nominal Ni valence from $Ni^{2.5+}$ to $Ni^{3+}$, reduces the Ni cation ionic radius, and consequently shortens the z-axis of the $NiO_6$ octahedra. The resulting crystal-field effect lifts the energy of the Ni-$3d_{z^2}$ orbital, manifesting as the marked expansion of the γ pocket, which renders Fermi-surface nesting less favorable. Figure 4b presents the electronic band structure at 25 GPa. Pressure broadens the bandwidth of the Ni-$e_g$ bands. The top of the bonding band becomes less flat than that in the ambient pressure. Moreover, the gap between the bonding and antibonding bands is enlarged. Wannier downfolding further reveals that the interlayer hopping of the Ni-$3d_{z^2}$ orbital increases from -0.551 eV at ambient pressure to -0.648 eV at 25 GPa, while the in-plane hopping of the Ni-$3d_{x^2-y^2}$ orbital increases from -0.473eV to -0.540 eV. The electron occupation of Ni-$3d_{z^2}$ and $3d_{x^2-y^2}$ orbitals are 0.92 and 1.08, respectively, with the $3d_{z^2}$ occupation being nearly halved compared to $La_3Ni_2O_7$[61]. In bilayer RP nickelates, superconductivity is frequently attributed to the interlayer superexchange coupling $J_{\perp}^{z}$[40]. An enhanced interlayer $3d_{z^2}$ hopping would strengthen $J_{\perp}^{z}$, but a reduced $3d_{z^2}$ occupation weakens it. Therefore, in $La_2SrNi_2O_7$, even though the interlayer $3d_{z^2}$ hopping under pressure can reach a magnitude comparable to that in high-pressure $La_3Ni_2O_7$, the introduction of excess holes suppresses the interlayer superexchange coupling and strongly quenches superconductivity.

Given these theoretical results, we propose two complementary strategies that target the key factors to potentially realize superconductivity. First, the excess holes could be compensated by simultaneously reducing the Sr doping level and substituting La with a higher-valence lanthanide ion such as Ce. The lower Sr content directly diminishes the hole concentration, while $Ce^{4+}$ donates electrons. This combined approach effectively increases the $3d_{z^2}$ occupation, which would recover the interlayer superexchange coupling $J_{\perp}^{z}$. Second, partial replacement of La by smaller lanthanide ions (e.g., $Sm^{3+}$ or $Nd^{3+}$) exerts chemical pressure and increases interlayer $3d_{z^2}$ hopping[62]. The combined approach may thus open a viable pathway to stabilize superconductivity in bilayer RP nickelates under ambient pressure.

## Conclusions

In summary, we have successfully synthesized $La_2SrNi_2O_{7-\delta}$ single crystals under HPHT conditions. At ambient pressure, the material adopts a tetragonal structure with undistorted $NiO_6$ octahedra, exhibiting metallic transport with a low-temperature upturn and no signatures of DW order. The tetragonal structure with the suppression of DW has previously been associated with high temperature superconductivity in bilayer nickelates under high pressure and compressive strain. However, despite systematic variations of pressure and oxygen content, no superconductivity is observed in our samples. Theoretical calculations reveal that holes are predominantly doped into the Ni-$d_{z^2}$ orbital, leading to a significantly enlarged γ Fermi pocket, whereas the α and β pockets stay nearly intact. The absence of superconductivity is attributed to a strong deviation from half-filling of the Ni-$d_{z^2}$ band, which is essential for interlayer antiferromagnetic coupling and electron pairing. This finding suggests that the precise position of the γ pocket, rather than its mere presence, is crucial for achieving superconductivity in this system.

## Experimental Section

### *Single crystal growth*

The precursor of $LaSrNiO_4$ and $LaNiO_3$ were synthesized via a modified sol-gel method. Stoichiometric amounts of $La_2O_3$ (99.99%, Alfa Aesar, dried at 1000°C prior to use), $SrCO_3$ (99.99% Alfa Aesar) and $Ni(OH)_2$ (99%, Picasso) were weighed and dissolved in nitric acid. The mixture was heated to 90 °C, then equimolar citric acid and ethylene glycol were added to achieve a homogeneous state. After heating for several hours, the green gel was formed and calcined at 500 °C for 10 h in a muffle furnace. After grinding, the dark grey product was further processed. $LaSrNiO_4$ was sintered at 1100 °C under 1 atm of flowing $O_2$ to promote densification and oxygen equilibration. $LaNiO_3$, on the other hand, was sintered at 750 °C under high oxygen pressure to stabilize its phase. Both materials were subjected to multiple sintering cycles to achieve a pure phase.
The synthesis of $La_2SrNi_2O_{7-\delta}$ was conducted under high-pressure and high-temperature (HPHT) conditions. Stoichiometric amounts of $LaSrNiO_4$ and $LaNiO_3$ were mixed in an equimolar ratio as the starting precursors. A eutectic $SrCl_2$ (99.5%, Alfa Aesar) and KCl (99.8%, Aladdin) mixture was employed as a flux to promote crystal growth and homogeneity. Additionally, $SrO_2$ was introduced into the reaction system to provide an extra oxygen source, thereby maintaining a sufficiently high oxygen partial pressure during the synthesis. The entire assembly was then subjected to a pressure of 3 GPa and heated to 1400 °C using a piston-cylinder-type high-pressure apparatus (LP 1000–540/50, Max Voggenreiter), leading to the formation of the desired $La_2SrNi_2O_{7-\delta}$ phase.

***Crystal Structural Determination and Sample Characterization***

The crystal structure of $La_2SrNi_2O_{7-\delta}$ was identified by single-crystal XRD (Bruker, D8 Venture PHOTON-II area detector) with Mo-Kα radiation (λ = 0.71073 Å). The data were collected at 193.15(10) K and the structure were solved and refined using Olex2 with ShelXT and ShelXL packages[63–65]. Powder X-ray diffraction (XRD) data were collected on a Bruker D8 Advance diffractometer equipped with Cu-Kα radiation (λ = 1.54184 Å) over the $2\theta$ range of 10°–120°. We used TOPAS 4.2 for Rietveld refinements[66], taking the single-crystal structure model as the starting model. Further details of the crystal structure of $La_2SrNi_2O_7$ can be found at the joint CCDC/FIZ Karlsruhe online deposition service under deposition number CSD 2574391. Scanning electron microscopy (SEM, Phenom ProX) provided single-crystal micrographs, while chemical compositions were determined using the energy-dispersive X-ray spectrometer (EDS) attached to the SEM. Thin-foil TEM specimens of single-crystal $La_2SrNi_2O_{7-\delta}$ were prepared via focused ion beam (FIB) milling on a Carl Zeiss Crossbeam 550L system. Characterization was subsequently performed on a Titan Themis Z microscope (Thermo Fisher Scientific) equipped with an aberration corrector, operated at 300 kV with a convergence angle of 25 mrad, with atomic-resolution images acquired separately in high-angle annular dark-field (HAADF) and integrated differential phase contrast (iDPC) modes. Nanoscale elemental maps were collected with the integrated Super-X energy-dispersive X-ray spectroscopy (EDS) system equipped with four windowless silicon-drift detectors, and atomic-scale elemental maps were obtained after post-processing with Gaussian blur and radial Wiener filtering in Velox software.

**Magnetization and resistivity measurements**

We carried out dc magnetization measurements using a superconducting quantum interference device (SQUID-VSM-7 T, Quantum Design). Temperature-dependent resistance measurements on polycrystalline pellets were performed with a physical property measurement system (PPMS-9 T, Quantum Design). The polycrystalline pellets were prepared by grinding microcrystals and pressing the resulting powder. Additionally, an annealing treatment at 500 °C under air or 1 atm $O_2$ was performed to enhance interparticle connectivity.

High-pressure transport measurements: Pressures were calibrated at room temperature using the ruby fluorescence method. A diamond anvil cell (DAC-PPMS-ET225, Shanghai Anvil Source Material Technology Co., Ltd) with a 300 μm culet was employed to generate pressures up to 52 GPa, and the four-probe van der Pauw method was applied for resistance measurements with KBr as the pressure-transmitting medium.

**Oxygen modification treatment**

*Reduction process*: The as grown crystals was ground and pressed into pellets. Then the pellet was sealed in an evacuated quartz tube, together with La metal. The sealed tube was heated to 330 °C for 5 h.

*Ozone annealing*: Ozone was generated from high-purity oxygen by means of an ozone generator, while a gas flow meter placed between the oxygen cylinder and the generator enabled control over the gas flow rate and the resulting ozone concentration delivered to the tube furnace. Ozone treatment was carried out at 4.2 wt% and 325 °C.

**DFT calculations**

DFT calculations were performed using Vienna ab initio simulation package implementing the projector-augmented wave method with a 550-eV plane-wave cut-off energy[67,68]. The generalized gradient approximation of Perdew-Burke-Ernzerhof form was used for exchange-correlation functional[69]. A 19 × 19 × 19 and 29 × 29 × 29 k-points mesh was used for the self-consistent and Fermi surface calculations. The lattice constants at ambient pressure were fixed to the structure identified by single-crystal XRD. High-pressure structural relaxations were performed starting from the ambient-pressure crystal structures after applying 4% strain. The inner atomic positions were fully optimized until forces on each atom were less than 0.001 eV $Å^{-1}$, and the energy convergence criterion was set at $10^{-7}$ eV for each electronic self-consistent loop. To handle the fractional atomic occupations of La and Sr, we used the virtual crystal approximation for electronic structure calculations and structural relaxations[70]. The XRD determined La: Sr ratio was employed. The results of different doping sites are provided in the Supplementary Materials, which shows negligible differences compared with those of the experimentally determined doping ratio. In all calculations, an effective Hubbard $U$ of 3.5 eV was applied for the 3$d$ electrons of Ni[39]. Wannier downfolding was performed using WANNIER90 package[71].

**Author Contributions**

The growth of $La_2SrNi_2O_{7-\delta}$ single crystals, SEM and EDS analyses were performed by Y.Z, X.M, and X.Z. The SXRD and PXRD data were collected by Y.Z, X.M, and X.Z. The resistivity and magnetization measurements at ambient pressure were done by Y.Z and X.M. The high-pressure electrical resistance measurements were conducted by Z.-N.X and Q.L. Theoretical calculations and analysis were finished by D.-X. Y and C.-Q. C. The STEM measurements were carried out by W.C. and B.-H.G. The manuscript was written by X.Z., Y.Z., X.M. D.-X.Y. and H.-H.W. All authors joined the analysis and agreed to publish the data.

**Acknowledgements**
This work was supported by the National Key Research and Development Program of China (No. 2022YFA1403201), National Natural Science Foundation of China (Nos. 52472276, 12434004, 11927809, 12504163, 12494591, 92565303), Natural Science Foundation of Jiangsu Province Grant (No. BK20233001), Natural Science Foundation of Shandong Province (Nos. ZR2025QC1499) and the Taishan Scholar Project of Shandong Province (No. tsqn202408178). We are grateful for the useful discussions with Kazuhiko Kuroki, Ilya Eremin and Hanghui Chen. We thank Zhenyi Zhang and Dongjing Hong for assistance with the single-crystal structure determination.

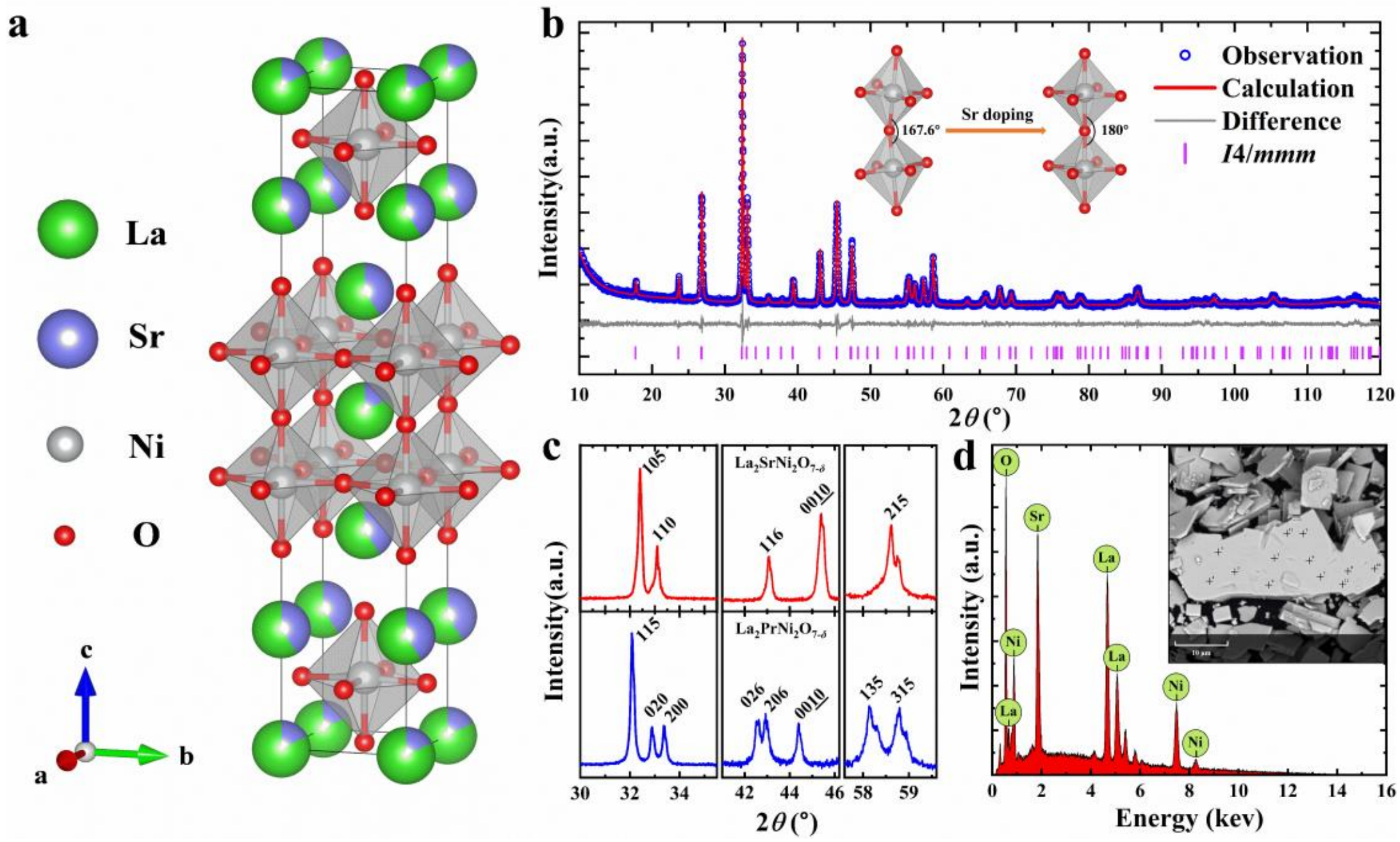

**Fig. 1| Characterizations of structure and composition of $La_2SrNi_2O_{7-\delta}$ single crystals.** (a) Crystal structure of $La_2SrNi_2O_{7-\delta}$ refined from SXRD data. (b) Rietveld refinements of PXRD pattern collected from crushed $La_2SrNi_2O_{7-\delta}$ single crystals. The inset illustrates the structural evolution of the $NiO_6$ octahedra upon heavy Sr-doping. (c) Detailed analysis of the enlarged PXRD patterns of $La_2SrNi_2O_{7-\delta}$, compared with orthorhombic $La_2PrNi_2O_{7-\delta}$. The clear adjacent-peak merging and (00*l*) peak shifts indicate the structural transition and a compression of the *c*-axis lattice parameter induced by Sr substitution. (d) EDS for the $La_2SrNi_2O_{7-\delta}$ single crystals. The right inset displays the SEM image of micro-crystals (scale bar: 10 μm).

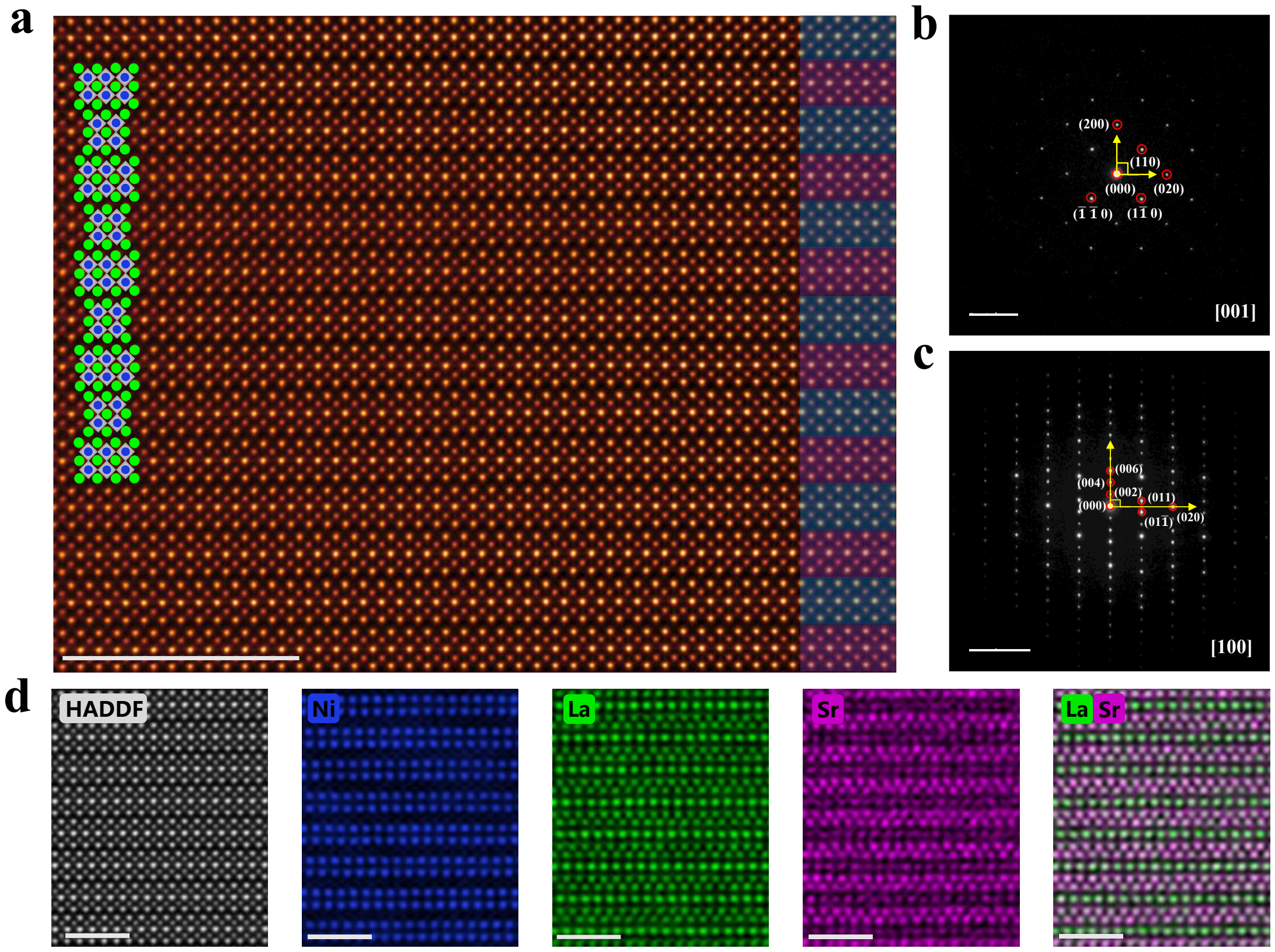

**Fig. 2| Atomic structure of the as-grown $La_2SrNi_2O_{7-\delta}$ single crystals.** (a) HAADF-STEM image of the $La_2SrNi_2O_{7-\delta}$ single crystal viewed along the [100] projection. (b, c) SAED patterns of the $La_2SrNi_2O_{7-\delta}$ single crystal along the (b): [001], (c): [100] axis. (d) HAADF image and corresponding atomic-resolution EDS elemental maps of La, Sr, and Ni acquired from the same region. Scale bars: 5 nm (a); 5 1/nm (b, c); 2 nm (d).

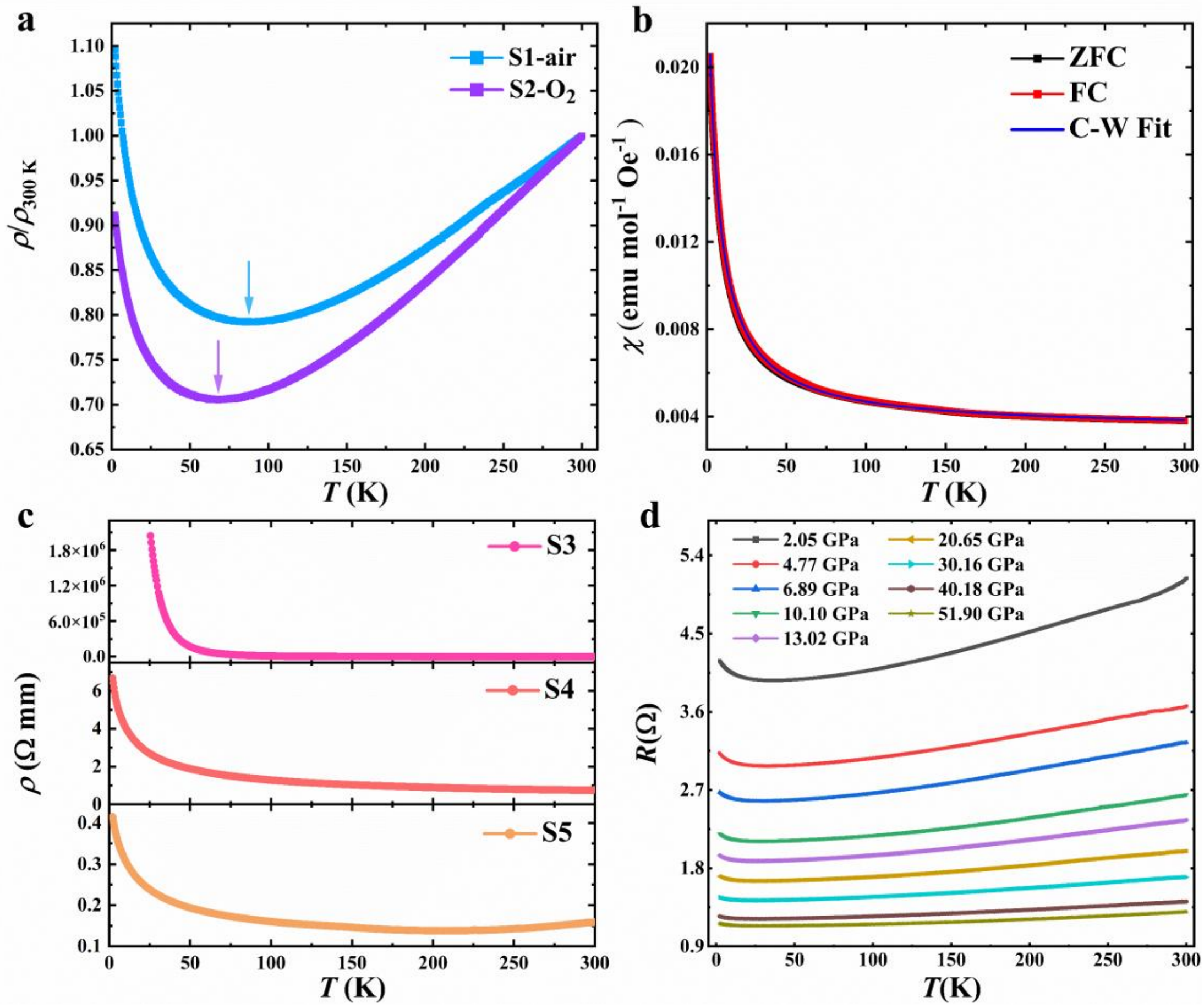

**Fig. 3| The electrical transport and magnetic properties of $La_2SrNi_2O_{7-\delta}$.**
(a) The normalized resistance of $La_2SrNi_2O_{7-\delta}$ annealed at 500 °C in air (S1) and oxygen flow (S2). (b) Temperature dependence of magnetic susceptibility measured at 5 kOe. The orange curve of $1/\chi$-$\chi_0$ versus $T$ exhibits a slightly nonlinearity transition consistent with transport measurements. (c) Transport results of $La_2SrNi_2O_{7-\delta}$ under different oxygen modification conditions: S3 was treated with La metal in a sealed quartz tube in order to reduce oxygen; S4 and S5 were then obtained by ozone annealing of S3 at 325 °C for 1 h and 3 h, respectively. (d) High-pressure electrical resistance measurements on $La_2SrNi_2O_{7-\delta}$ single crystals in a diamond-anvil-cell (DAC) apparatus.

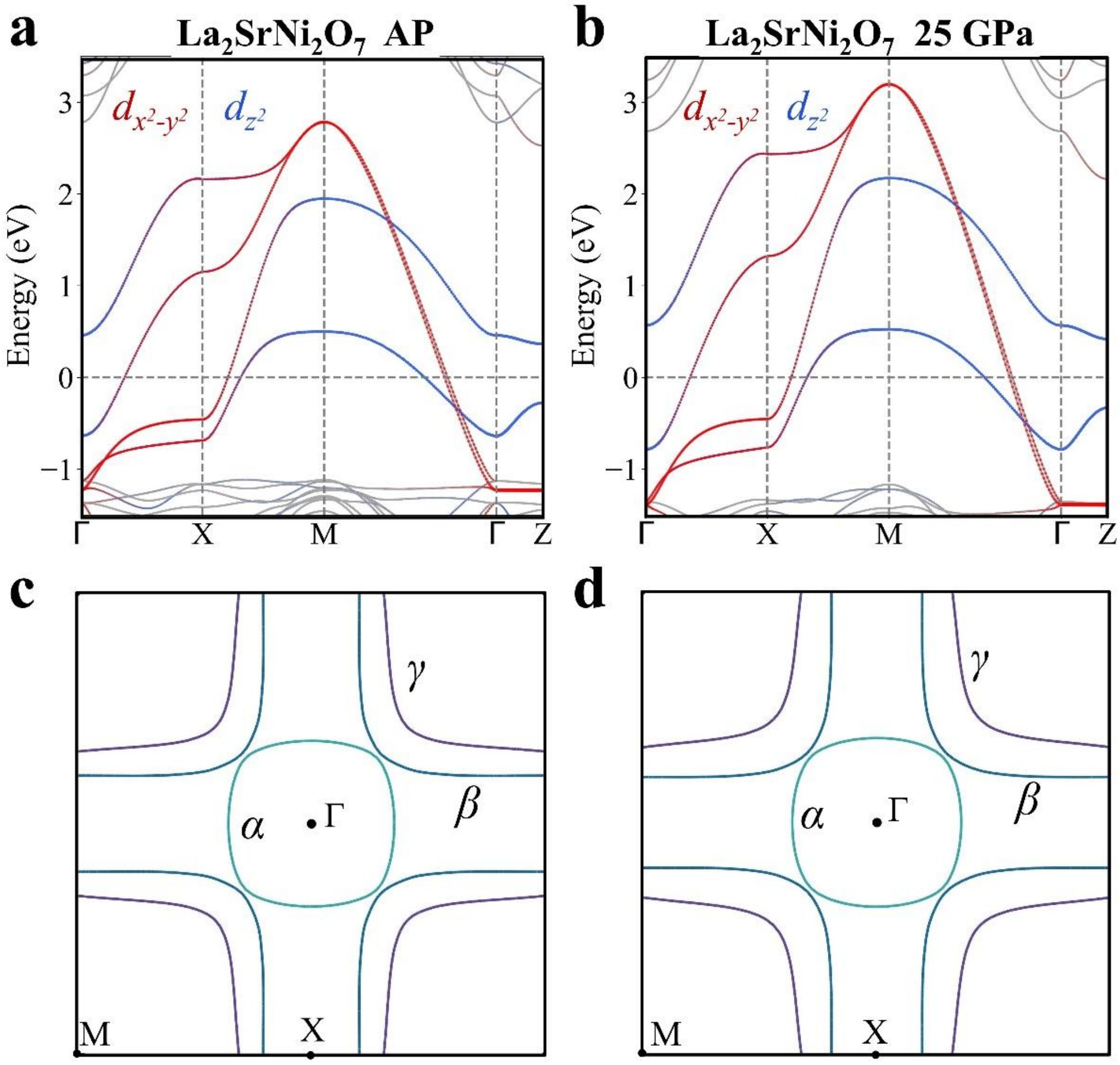

**Fig. 4| DFT calculations for $La_2SrNi_2O_{7-\delta}$ at ambient pressure and 25 GPa.** (a)-(b) Projected band structures of Ni-$e_g$ orbitals in $La_2SrNi_2O_7$ at ambient pressure (a) and 25 GPa (b). The red and blue colors represent the contributions from Ni-3$d_{z^2}$ and Ni-3$d_{x^2-y^2}$ orbitals. (c)-(d) Calculated two-dimensional Fermi surface at $k_z$ = 0 slice at ambient pressure (c) and 25 GPa (d). Different colors denote distinct band indices.

**Table 1. Crystallographic data and structure refinement for $La_2SrNi_2O_{7-\delta}$ obtained from the SXRD**

| Empirical formula | $La_2SrNi_2O_{7-\delta}$ | | | | |
|---|---|---|---|---|---|
| Crystal system | tetragonal | | | | |
| Space group | *I 4/m m m* | | | | |
| Temperature [K] | 193.2 | | | | |
| Formula weight | 594.86 | | | | |
| a [Å] | 3.8264(2) | | | | |
| b [Å] | 3.8264(2) | | | | |
| c [Å] | 19.9908(17) | | | | |
| α [°] | 90 | | | | |
| β [°] | 90 | | | | |
| γ [°] | 90 | | | | |
| Volume [$Å^3$] | 292.69(4) | | | | |
| Density (calculated) [g $cm^{-3}$] | 6.750 | | | | |
| Z | 2 | | | | |
| Radiation type | Mo-$K_\alpha$ (λ =0.71073 Å) | | | | |
| Crystal size [$mm^3$] | 0.047×0.045×0.002 | | | | |
| Absorption coefficient [$mm^{-1}$] | 29.634 | | | | |
| Data collection diffractometer | Bruker D8 VENTURE dual source, PHOTON-III detector | | | | |
| Absorption correction | multi-scan | | | | |
| Reflections collected | 4015 | | | | |
| Independent reflections | 175 ($R_{int}$ = 0.0680) | | | | |
| θ range for data collection [°] | 2.037-30.529 | | | | |
| *F* (000) | 528 | | | | |
| Index ranges | -5≤h≤5, -5≤k≤5, -26≤l≤28 | | | | |
| Data, restraints, parameters | 175, 79, 20 | | | | |
| Goodness of fit on $F^2$ | 1.097 | | | | |
| Final *R* indexes (I > 2σ) | $R_1$= 0.0219, $wR_2$= 0.0549 | | | | |
| Final *R* indexes (all data) | $R_1$= 0.0299, $wR_2$= 0.0612 | | | | |
| Largest diff. peak/hole [e $Å^{-3}$] | 2.88/-1.49 | | | | |
| **Site label** | ***x*** | ***y*** | ***z*** | **Occ.** | $U_{(eq)}$ |
| La1 | 0.5 | 0.5 | 0.5 | 0.824(10) | 0.0111(3) |
| Sr1 | 0.5 | 0.5 | 0.5 | 0.176(10) | 0.0111(3) |
| La2 | 0 | 0 | 0.18025(4) | 0.588(5) | 0.0104(3) |
| Sr2 | 0 | 0 | 0.18025(4) | 0.412(5) | 0.0104(3) |
| Ni1 | 0 | 0 | 0.40229(7) | 1 | 0.0065(3) |
| O1 | 0.5 | 0 | 0.4058(3) | 1 | 0.0135(12) |
| O2 | 0 | 0 | 0.5 | 1 | 0.017(2) |
| O3 | 0 | 0 | 0.3026(5) | 1 | 0.0155(18) |

The equivalent isotropic ($U_{eq}$) displacement parameter is defined as 1/3 of the trace of the orthogonalized $U_{ij}$ tensor. The unit of $U_{eq}$ is 1 $Å^2$.